# RLS-Based Adaptive Dereverberation Tracing Abrupt Position Change of Target Speaker

Teng Xiang
Key Laboratory of Modern Acoustics and
Institute of Acoustics
Nanjing University
Nanjing, China
txiang@smail.nju.edu.cn

Jing Lu
Key Laboratory of Modern Acoustics and
Institute of Acoustics
Nanjing University
Nanjing, China
lujing@nju.edu.cn

Kai Chen
Key Laboratory of Modern Acoustics and
Institute of Acoustics
Nanjing University
Nanjing, China
chenkai@nju.edu.cn

*Abstract*—Adaptive algorithm based on multi-channel linear prediction is an effective dereverberation method balancing well between the attenuation of the long-term reverberation and the dereverberated speech quality. However, the abrupt change of the speech source position, usually caused by the shift of the speakers, forms an obstacle to the adaptive algorithm and makes it difficult to guarantee both the fast convergence speed and the optimal steady-state behavior. In this paper, the RLS-based adaptive multi-channel linear prediction method is investigated and a time-varying forgetting factor based on the relative weighted change of the adaptive filter coefficients is proposed to effectively tracing the abrupt change of the target speaker position. The advantages of the proposed scheme are demonstrated in the simulations and experiments.

## I. Introduction

Dereverberation is aimed at reducing the reverberant part of speech to improve the speech quality for communication and decrease the word error rate (WER) of speech recognition systems. The directivity of the microphone array, realized through beamforming [1-2], can be utilized to capture more direct sound from the speech source and suppress the reverberation, but it usually requires a large number of microphones to obtain satisfactory spatial beampatterns. The spectral subtraction method is another possible approach in which the late reverberation is suppressed in the magnitude spectrum [3-4]. However, spectral subtraction usually suffers from significant speech distortion caused by phase, cross-term and magnitude errors [5]. Multi-channel linear prediction algorithm (MCLP), a kind of popular blind dereverberation method based on linear prediction model of speech [6-8], is able to preserve the speech quality while attenuating the reverberation effectively.

To overcome the defects of whitening effect of the conventional MCLP [9], the prediction delay was proposed to estimate the late reverberation, resulting in the delayed linear prediction (DLP) algorithm [10]. To further improve the dereverberation performance, the weighted prediction error (WPE) method was proposed, taking into account the time-varying speech characteristics [11]. The WPE algorithm can be conducted either in the time domain or in the short-time Fourier transform (STFT) domain. However, its batch processing makes it unsuitable for most time-critical applications and varying room impulse responses (RIRs) conditions. The online adaptive filter estimation methods were proposed in [12-14], where the recursive least squares (RLS) algorithm or the Kalman filter was adopted. In addition, the equivalence between RLS and Kalman filter has been elucidated in [15]. An improved RLS algorithm constraining the power of the late reverberation was proposed in [13] and has been validated with continual changes of the speaker positon. However, the fixed forgetting factor of the RLS algorithm makes it hard to effectively trace the abrupt change of speaker positions, a common scenario in practical applications usually caused by the shift of speakers.

In this paper, the tracking ability of the RLS algorithm for the MCLP method is analyzed intuitively. The relative change of the adaptive filter coefficients is shown to be sensitive to the abrupt change of the speaker positon, and it can be carefully regularized to optimize a time-varying forgetting factor to improve the tracking ability of the online adaptive algorithm. The benefits of the proposed method are validated in the following simulations and experiments.

## II. RLS-based MCLP Method

Consider an acoustic scenario where a single speech source is captured by $M$ microphones in a reverberant room without noise. Let $x_k^{(m)}(n)$ denote the observed signal acquired by the $m$th microphone at frame $n$ and frequency bin $k$ in the STFT domain. It can be decomposed into the desired signal $d_k^{(m)}(n)$ and the late reverberation $r_k^{(m)}(n)$ as

$$x_k^{(m)}(n) = d_k^{(m)}(n) + r_k^{(m)}(n), \qquad (1)$$

where $d_k^{(m)}(n)$ contains the direct signal and the early reflections. The late reverberation can be estimated from the observed signal by using the MCLP method so that the desired signal can be obtained as

$$d_k^{(m)}(n) = x_k^{(m)}(n) - \mathbf{w}_k^\mathrm{T}(n)\mathbf{x}_k(n-D), \qquad (2)$$

where $\mathbf{w}_k(n)$ is the frame dependent filter coefficients, $D$ is the prediction delay, the superscript $(\cdot)^\mathrm{T}$ denotes the transpose operation, and

This work was supported by the National Natural Science Foundation of China (Grant No. 11374156).

$$\mathbf{x}_k(n) = \left[\underline{\mathbf{x}}_k^T(n), \underline{\mathbf{x}}_k^T(n-1), \cdots, \underline{\mathbf{x}}_k^T(n-L_w+1)\right]^T, \quad (3)$$

with $\underline{\mathbf{x}}_k(n) = [x_k^{(1)}(n), x_k^{(2)}(n), \ldots, x_k^{(M)}(n)]^T$ and $L_w$ the prediction order.

The cost function based on the weighted least squares principle can be defined as [12]

$$\mathcal{J}(\mathbf{w}_k(n)) = \sum_{\tau=1}^{n} \lambda^{n-\tau} \frac{\left|d_k^{(m)}(\tau)\right|^2}{\sigma_k^2(\tau)} + \alpha\lambda^n \left\|\mathbf{w}_k(n)\right\|_2^2, \quad (4)$$

where $\sigma_k^2(\tau)$ is the variance of the desired signal, $\lambda$ is the forgetting factor, $\alpha$ is a regularization factor, and $\|\cdot\|_2$ denotes the $l_2$ norm. The closed form solution of the adaptive filter coefficients can be expressed as

$$\hat{\mathbf{w}}_k(n) = \boldsymbol{\Psi}_k^{-1}(n)\mathbf{z}_k(n), \quad (5)$$

where

$$\boldsymbol{\Psi}_k(n) = \sum_{\tau=1}^{n} \lambda^{(n-\tau)} \frac{\mathbf{x}_k^*(\tau-D)\mathbf{x}_k^T(\tau-D)}{\sigma_k^2(\tau)} + \alpha\lambda^n \mathbf{I} \quad (6)$$

and

$$\mathbf{z}_k(n) = \sum_{\tau=1}^{n} \lambda^{n-\tau} \frac{\mathbf{x}_k^*(\tau-D)x_k^{(m)}(\tau)}{\sigma_k^2(\tau)}, \quad (7)$$

the superscript $(\cdot)^*$ denotes the complex conjugation and $\mathbf{I}$ is the identity matrix. From (5) – (7), the RLS algorithm can be obtained by using the matrix inversion lemma, and the updating processes are as follows.

$$\mathbf{k}_k(n) = \frac{\lambda^{-1}\mathbf{P}_k(n-1)\mathbf{x}_k^*(n-D)}{\sigma_k^2(n) + \lambda^{-1}\mathbf{x}_k^T(n-D)\mathbf{P}_k(n-1)\mathbf{x}_k^*(n-D)}, \quad (8)$$

$$d_k^{(1)}(n) = x_k^{(1)}(n) - \hat{\mathbf{w}}_k(n-1)^T \mathbf{x}_k(n-D), \quad (9)$$

$$\hat{\mathbf{w}}_k(n) = \hat{\mathbf{w}}_k(n-1) + \mathbf{k}_k(n)d_k^{(1)}(n), \quad (10)$$

and $\mathbf{P}_k(n) = \lambda^{-1}\mathbf{P}_k(n-1) - \lambda^{-1}\mathbf{k}_k(n)\mathbf{x}_k^T(n-D)\mathbf{P}_k(n-1), \quad (11)$

where $\mathbf{P}_k(n)$ denotes the inversion of matrix $\boldsymbol{\Psi}_k(n)$.

The ML-based estimation proposed in [16] can be used to estimate the variance $\sigma_k^2(n)$ of the desired signal, which may degrade the performance of the RLS algorithm when the speaker position changes. In this paper, the variance is estimated as

$$\sigma_k^2(n) = \beta\sigma_k^2(n-1) + (1-\beta)\left|x_k^{(1)}(n)\right|^2, \quad (12)$$

where $\beta$ is a recursive smoothing factor. This simple and straightforward estimation shows high robustness in our numerous experiments in adverse situations.

## III. PROPOSED METHOD

### A. Tracking Ability of the RLS-based Method

Suppose that the acoustic system can be modeled by a first-order Markov model as

$$\mathbf{w}_{0,k}(n) = \mathbf{w}_{0,k}(n-1) + \boldsymbol{\omega}_k(n), \quad (13)$$

where $\mathbf{w}_{0,k}(n)$ is the perfect filter coefficients and $\boldsymbol{\omega}_k(n)$ is the process noise. The transition matrix is assumed to be an identity matrix. When $n$ is sufficiently large, i.e. the RLS algorithm has converged, the deviation of the estimated filter coefficients can be approximately estimated by [17]

$$\mathrm{E}\left[\left\|\boldsymbol{\varepsilon}_k(n)\right\|_2^2\right] \approx \frac{1-\lambda}{2}\mathrm{tr}\left[\mathbf{R}_{\mathbf{x},k}^{-1}\right] + \frac{1}{2(1-\lambda)}\mathrm{tr}\left[\mathbf{R}_{\boldsymbol{\omega},k}\right], \quad (14)$$

where

$$\boldsymbol{\varepsilon}_k(n) = \mathbf{w}_{0,k}(n) - \hat{\mathbf{w}}_k(n), \quad (15)$$

$$\mathbf{R}_{\mathbf{x},k} = \mathrm{E}\left[\frac{\mathbf{x}_k^*(n-D)\mathbf{x}_k^T(n-D)}{\sigma_k^2(n)}\right] \quad (16)$$

and

$$\mathbf{R}_{\boldsymbol{\omega},k} = \mathrm{E}\left[\boldsymbol{\omega}_k(n)\boldsymbol{\omega}_k^H(n)\right]. \quad (17)$$

If the speaker stands still, the process noise is very small and the second term on the right side of (14) can be neglected. Therefore, the forgetting factor should be set close to 1 to insure a good steady-state behavior. When the speaker position changes, the RLS algorithm can be regarded as being reinitialized. According to (5) and (6), a small forgetting factor is needed for the algorithm to swiftly trace the variation of the matrix $\boldsymbol{\Psi}_k(n)$.

### B. Relative Change of the Filter Coefficients

The relative change of the filter coefficients at each frequency bin is defined as

$$\delta_k(n) = \frac{\left\|\hat{\mathbf{w}}_k(n) - \hat{\mathbf{w}}_k(n-1)\right\|_2^2}{\left\|\hat{\mathbf{w}}_k(n)\right\|_2^2}, \quad (18)$$

and the total relative change is the sum over all the relative changes as

$$\delta_{\text{T}}(n) = \beta_w \delta_{\text{T}}(n-1) + (1-\beta_w)\sum_{k=0}^{K-1}\delta_k(n), \qquad (19)$$

where $\beta_w$ is a recursive smoothing factor. It is intuitive that this total relative change closely relates to the convergence state of the system and is expected to reflect the abrupt change of the speaker positions. However, as illustrated in Fig. 1, the variation of the filter coefficients caused by the abrupt change of the speaker position is highly likely to be masked by the variation caused by the fluctuation of the speech amplitude. Therefore, a factor $\rho_k(n)$ related to the power of the speech signal is further utilized to obtain the weighted relative change as

$$\delta_{\text{w},k}(n) = \frac{\rho_k(n)\|\hat{\mathbf{w}}_k(n)-\hat{\mathbf{w}}_k(n-1)\|_2^2}{\|\hat{\mathbf{w}}_k(n)\|_2^2}, \qquad (20)$$

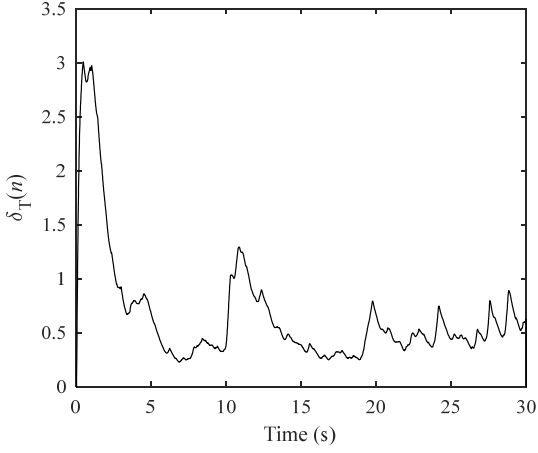

Fig. 1. Total relative change of the filter coefficients estimated by (19) with an abrupt change of speaker position at 10 s ($\lambda = 0.998$).

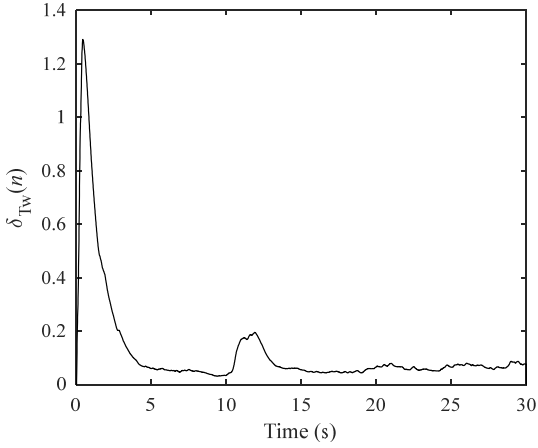

Fig. 2. Total weighted relative change of the filter coefficients estimated by (21) with an abrupt change of speaker position at 10 s ($\lambda = 0.998$).

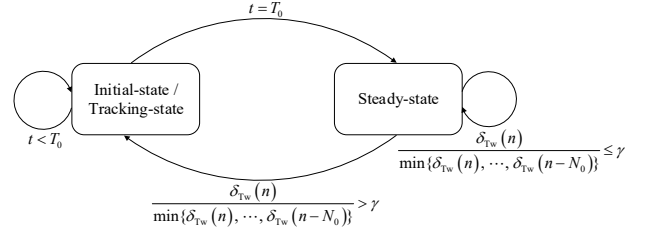

Fig. 3. The finite-state machine controlling the forgetting factor for RLS-based adaptive MCLP dereverberation method.

where the weighting factor $\rho_k(n)$ is set as the square root of the variance of the desired signal estimated by (12). The total weighted relative change is smoothly estimated as

$$\delta_{\text{Tw}}(n) = \beta_w \delta_{\text{Tw}}(n-1) + (1-\beta_w)\sum_{k=0}^{K-1}\delta_{\text{w},k}(n). \qquad (21)$$

Fig. 2 presents the total weighted relative change of the filter coefficients estimated by (21), where a sudden increase of the total weighted relative change can be clearly seen at around 10 s, the time when a sudden change of the speaker position occurs.

### C. Time-varying Forgetting Factor

According to the analyses in Secs. III. A and B, a time varying forgetting factor is designed based on the total weighted relative change of the filter coefficients. The abrupt change of the speaker position can be inferred using the following criterion

$$\frac{\delta_{\text{Tw}}(n)}{\min\{\delta_{\text{Tw}}(n),\cdots,\delta_{\text{Tw}}(n-N_0)\}} > \gamma, \qquad (22)$$

where $\gamma$ is a magnitude threshold and $N_0$ is a time threshold. A smaller forgetting factor $\lambda_S$ should be set at the initialization of the system or when the speaker position changes, so that a fast convergence speed can be achieved. After a fixed short time $T_0$, a higher forgetting factor $\lambda_L$ needs to be set for a better steady-state performance, and $\lambda_L$ varies as

$$\lambda_L = \max\left\{2-e^{\epsilon \delta_{\text{Tw}}(n)}, \lambda_{L0}\right\}, \qquad (23)$$

where $\lambda_{L0}$ is a lower limit of $\lambda_L$ and $\epsilon$ is a scale factor. A finite-state machine controlling the variation of the forgetting factor is shown in Fig. 3, where $t$ is a timer that is initialized as 0 and is reset to 0 when the speaker position changes.

## IV. SIMULATIONS AND EXPERIMENTS

### A. Configurations

In all our simulations and experiments, the sampling frequency is 16 kHz, the block size of STFT is 512 points with 128 points shift between adjacent blocks, and a Hanning window is used. The prediction delay and the prediction order are set at 2 and 40, respectively. The matrix $\mathbf{P}_k(n)$ is initialized by a scaled identity matrix $\alpha^{-1}\mathbf{I}$ with $\alpha$ equal to 100. At the initial/tracing state, the fixed time threshold $T_0$ is 375 points and the forgetting factor $\lambda_S$ is 0.990. The magnitude threshold $\gamma$ and time threshold $N_0$ for the determination of the abrupt change of the speaker position are set at 1.5 and 35, respectively. At the steady-state, $\lambda_{L0}$ is 0.998 and $\epsilon$ is 0.01. The smoothing factors $\beta$ and $\beta_w$ are 0.6 and 0.99, respectively. The above parameters shown in (22) - (23) are tweaked carefully and our numerous simulations and experiments validate their robustness in different reverberant environments.

The performance of the algorithms is evaluated with two objective measures, the perceptual evaluation of speech quality (PESQ) [18] and the short-time objective intelligibility (STOI) [19]. All of the objective scores are averaged over all the 32 different utterances in the simulations or the 64 recording sample sets in the experiments. The instantaneous STOI scores and the PESQ scores are computed with windows of 2 s and 75% overlap. Exemplary audio samples are available online at https://github.com/helianvine/RLS-based-MCLP.

*B. Simulations*

The image method [20] was used to simulate the reverberation in a room of size 6.5 × 5.1 × 3.8 m with different reverberation time. Three microphones were placed linearly with 0.05 m intervals in the room and the distance between the speaker and the center microphone was 2.2 m. Reverberant speech signals were synthesized by convolving the clean speech signals with the RIRs with reverberation time $T_{60}$ of 0.5 and 1.0 s respectively. The speaker position changed from 25° right to 25° left of the broadside direction of the microphone array at 10 s.

Fig. 4 shows the STOI scores for the observed reverberant speech signals ("REV") and dereverberated speech signals with different forgetting factors. As expected from the basic features of the RLS algorithm, the smaller fixed forgetting factor leads to a faster convergence speed while the larger one leads to a better steady-state behavior. Our proposed time-varying forgetting factor shows a significantly better balance between the convergence speed and the steady-state behavior. Note that a larger fixed forgetting factor leads to a so slow converge speed that the RLS algorithm cannot converge in short time periods, resulting in worse performance than that brought by the varied forgetting factor in most time shown in Fig. 4.

*C. Experiments*

A linear array of three microphones with 0.06 m intervals was used to capture the reverberant signals in a reverberant room with reverberation time of around 1.2 s. The speaker-microphone distance was 1.5 m. Two loudspeakers were located at around 30° right and 30° left of the broadside direction, respectively. The speech signals were firstly played by the left loudspeaker and switched to the right one at 10 s. An interfering signal with signal to interference ratio (SIR) of 20 dB was added when recording the reverberant signals to validate the performance of the whole system in a more complex environment.

The STOI and PESQ variations of the experimental results are depicted in Fig. 5. Similar to that demonstrated in the simulations, the proposed time-varying forgetting factor results in both faster convergence speed and optimal steady-state behavior.

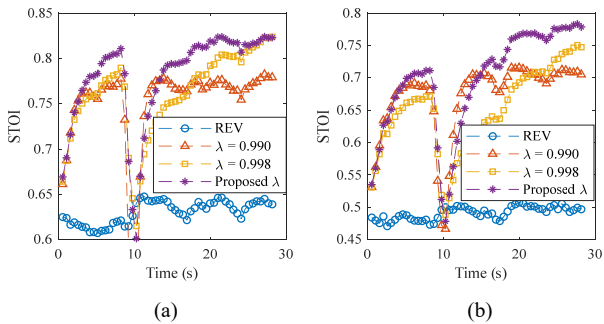

Fig. 4. STOI scores in simulations for the reverberant and dereverberated speech with different forgetting factor settings. (a) $T_{60}$ = 0.5 s. (b) $T_{60}$ = 1.0 s.

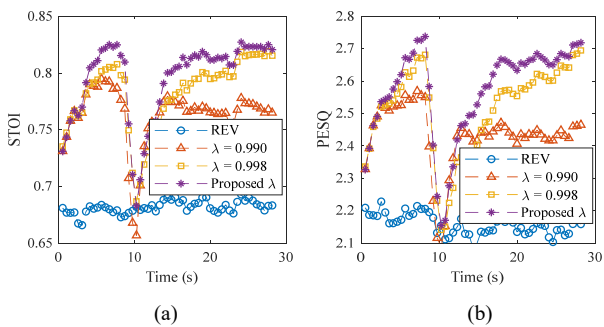

Fig. 5. Objective scores in experiments for the reverberant and dereverberated speech with different forgetting factor settings. (a) STOI curves. (b) PESQ curves.

V. CONCLUSIONS

In this paper, a time-varying forgetting factor is proposed to improve the RLS-based MCLP method for tracing the abrupt change of the target speaker position. The total weighted relative change is utilized to optimize the forgetting factor, leading to a fast convergence speed when the speaker position changes and the optimal attenuation of the long-term reverberation in steady-state situations. The efficacy of the proposed scheme is proven by the simulations with different reverberation time and the experiments in a highly reverberant environment. The comparison between the proposed algorithm and the convex combination of adaptive filters will be carried out in the future.